\documentclass[preprint2]{aastex}

\newcommand{\be}{\begin{equation}}
\newcommand{\ee}{\end{equation}} 

\shorttitle{Diffusive Acceleration}
\shortauthors{Fatuzzo and Melia}

\begin{document}

\title{Assessing the Feasibility of Cosmic-Ray Acceleration \\ by Magnetic Turbulence at the Galactic Center}

\author{M. Fatuzzo}
\affil{Physics Department, Xavier University, Cincinnati, OH 45207}
\email{fatuzzo@xavier.edu}

\and

\author{F. Melia}
\affil{Department of Physics, The Applied Math Program, and Steward Observatory, \\ 
The University of Arizona, AZ 85721}
\email{fmelia@email.arizona.edu}

\begin{abstract}
The presence of relativistic particles at the center of our galaxy 
is evidenced by the  diffuse TeV emission detected from the inner 
$\sim$$2^\circ$ of the Galaxy. Although it is not yet entirely clear 
whether the origin of the TeV photons is due to hadronic or leptonic 
interactions, the tight correlation of the intensity distribution 
with the distribution of molecular gas along the Galactic ridge
strongly points to a pionic-decay process involving relativistic 
protons. In earlier work, we  concluded that point-source candidates, 
such as the supermassive black hole Sagittarius A* (identified with
the HESS source J1745-290), or the pulsar wind nebulae dispersed
along the Galactic plane, could not account for the observed diffuse 
TeV emission from this region. Motivated by this result, we consider 
here the feasibility that the cosmic rays populating the Galactic 
Center (GC) region are accelerated in situ by magnetic turbulence. Our 
results indicate that even in a highly conductive environment, this 
mechanism is efficient enough to energize protons within the intercloud 
medium to the $\ga $ TeV energies required to produce the HESS emission. 

\end{abstract}


\section{Introduction}

Observations of the Galactic Center (GC) with the High-Energy Stereoscopic
System (HESS) have revealed the presence of  diffuse TeV emission
spread out roughly $0.2^\circ$ in Galactic latitude $b$ within the inner $2^\circ$ of our galaxy
(Aharonian et al. 2006).
The strong correlation between the $> 200$ GeV emission and the 
$\sim$$10^8\;M_\odot$ of molecular gas distributed along the GC ridge,
as traced
by its CO and CS line emission (see, e.g., Tsuboi et al. 1999),
points to the decay of neutral pions produced by the scattering of relativistic cosmic rays
with the proton-rich target of overlapping clouds as the dominant source of this diffuse radiation
 (see, e.g., Crocker et al. 2005; Ballantyne et al. 2007).

The origin of these energetic hadrons is an intriguing puzzle because
the observed gamma-ray spectrum requires an underlying
cosmic-ray population quite different from that seen at Earth. 
Specifically, the gamma-ray spectrum measured by HESS in the
region $|l|<0.8^\circ$ and $|b|<0.3^\circ$ (with point-source emission subtracted)
can be reasonably fit with a power law with photon index $\Gamma=2.29\pm 0.27$.
Since the spectral index of the gamma rays tracks the spectral index of the 
cosmic rays themselves, the implied cosmic ray index ($\sim$$2.3$) is then much 
harder than that ($\sim$$2.75$) measured locally.

Possible sources of energetic
hadrons at the GC were recently considered by Wommer et al. (2008).
The results of this effort seemingly rule out point sources such as 
Sagittarius A* and the pulsar wind nebulae dispersed along the 
Galactic ridge, and thereby give credence to the possibility that the 
relativistic protons are accelerated throughout the GC medium.
Following up on this result, Fatuzzo \& Melia (2011) found that stochastic
acceleration by magnetic turbulence within the intercloud medium,
which is effectively a 1-D random walk in energy process, 
can produce a distribution of particle 
energies whose high-energy tail is capable of reproducing the HESS data,
so long as the tail extends to $\ga 1$ TeV energies.  

However, the electric field used to calculate the energy evolution 
of particles in Fatuzzo \& Melia (2011), which  was derived directly
from the form of the turbulent magnetic field using Faraday's Law, 
had a component parallel to the overall magnetic field.  As a result, 
the electric field was so efficient at energizing protons that 
the required particle distributions  could only be produced if the turbulent field was
much weaker than the underlying field or if acceleration was  
limited to small ``active regions" within the intercloud medium. 
However, the high conductivity of the medium within the GC environment
makes it highly likely that a component of the electric field 
parallel to the underlying magnetic field would be quickly ``quenched".

The purpose of this paper is to reassess the feasibility of stochastic acceleration
within the GC region with the (more realistic) assumption that the highly conductive medium
does not allow for an electric field component parallel to the magnetic field. 
Toward that end, we adopt the formalism of O'Sullivan et al. (2009) to construct the
turbulent fields so that the electric field is everywhere perpendicular to the total magnetic field.
The spatial and energy diffusion of cosmic-ray protons within the molecular
cloud and intercloud regions are then investigated via numerical simulations.  
Specifically, both the spatial 
and energy  diffusion coefficients
over a relevant range of parameter space are calculated and used to 
compare estimates of the time required to energize protons up to TeV energies 
with the escape and cooling times associated with both the cloud and intercloud environments.

Our results indicate that protons in the intercloud medium can be energized up to the 
$\ga$ TeV energies required to produce the observed HESS emission.  As such,
stochastic particle acceleration by magnetic turbulence appears to be a viable mechanism
for Cosmic-ray production at the GC.  

\section[]{The Physical Conditions}
The large concentration (up to $\sim$$10^8\;M_\odot$) of dense molecular
gas at the GC is largely confined to GMC's with a size
$\sim$$50$--$70$ pc (G\"{u}sten \& Philipp 2004). These clouds appear to
be clumpy with high-density ($\sim$$10^5$ cm$^{-3}$) regions embedded
within less dense ($\sim$$10^{3.7}$ cm$^{-3}$) envelopes (e.g., Walmsley et al.
1986) and are threaded by a pervasive magnetic
field whose milligauss strength is suggested by the rigidity of non-thermal
filaments interacting with the molecular clouds  (see, e.g., 
Yusef-Zadeh \& Morris 1987; Morris \& Yusef-Zadeh 1989; Morris 2007). 
Confirming evidence for such field strengths in and around the GMC's
is provided by their apparent stability.
The observed pressure $P_{plasma}$$\sim$$10^{-9.2} \;{\rm erg}\;{\rm cm}^{-3}$
due to the hot plasma between the clouds is an order of magnitude
smaller than the turbulent pressure $P_{turb}$$\sim$$10^{-8} \;{\rm erg}\;{\rm cm}^{-3}$
within the GMC environment (G\"{u}sten \& Philipp 2004), 
seemingly ruling out pressure confinement.
If clouds are instead bound by their own magnetic fields, then
equating the turbulent and magnetic ($B^2/8\pi$) energy densities gives
field strengths of $\sim$$0.5$ mG within the clouds, not too different
from the typical value measured in the non-thermal filaments.

We are now also reasonably sure of the magnetic field strength between
the clouds. In the past, the field intensity near the GC 
had been uncertain by two orders of magnitude. We've just seen how on
a scale of $\sim$100 pc, field strengths can be as high as
$\sim$1 mG. At the other extreme,
equipartition arguments based on radio observations favor fields of only
$\sim$6 $\mu$G on $\sim$400 pc scales (LaRosa et al. 2005). But a more
careful analysis of the diffuse emission from the central bulge has
revealed a down-break in its non-thermal radio spectrum, attributable
to a transition from bremsstrahlung to synchrotron cooling of the in
situ cosmic-ray electron population. Crocker et al. (2010) have shown
recently that this spectral break requires a field of $\sim$50 $\mu$G
extending over several hundred parsecs, lest the synchrotron-emitting
electrons produce too much $\gamma$-ray emission given existing
constraints (Hunter et al. 1997).

While the structure of this magnetic field is not well understood,
magnetic fluctuations are expected to be present in essentially all 
regions of the interstellar medium. 
For example, molecular clouds are observed to have substantial
non-thermal contributions to the observed molecular line-widths (e.g.,
Larson 1981; Myers, Ladd, \& Fuller 1991; Myers \& Gammie 1999).
These non-thermal motions are generally interpreted as arising from
MHD turbulence (e.g., Arons \& Max 1975; Gammie \& Ostriker 1996; for
further evidence that the observed linewidths are magnetic in origin,
see Mouschovias \& Psaltis 1995).  Indeed, the size of these
non-thermal motions, as indicated by the observed line-widths, are
consistent with the magnitude of the Alfv{\'e}n speed (e.g., Myers \&
Goodman 1988; Crutcher 1998, 1999; McKee \& Zweibel 1995; Fatuzzo \&
Adams 1993). As a result, the fluctuations are often comparable in
magnitude to the mean values of the fields.

For this work, we treat molecular clouds within the GC 
as spherical ($R_c = 30$ pc), uniform density ($n_H = n_{H_2}/2 
= 10^4$ cm$^{-3}$) structures threaded by an underlying 
uniform magnetic field $\vec B_0 = B_0 \hat z$, where
$B_0 = 0.50$ mG.  The Alfv\'en speed within the cloud
environment is therefore taken to be $v_A \approx 11$ km/s.
Consistent with the limits placed by Crocker et al. (2010), 
we treat the intercloud medium as a spherical ($R_{ic} = 200$ 
pc),  low density ($n_H = 10$ cm$^{-3}$) structure threaded by an 
underlying uniform magnetic field $\vec B_0 = B_0 \hat z$, where
$B_0 = 50\,\mu$G.  The Alfv\'en speed within the intercloud  
environment is therefore $v_A = 35$ km/s.  For both regions,
we assume that the magnetic turbulence has the same energy density
as that of the underlying uniform magnetic field.  

For completeness, we note that the
conditions much closer to Sagittarius A* are somewhat different and
appear to be controlled primarily by ongoing stellar wind activity
(Rockefeller et al. 2004). But this is a very small region compared
to the rest of the TeV emitting gas, so we do not expect it to
significantly influence our results.
 
\section{The Turbulent Fields}

The standard numerical approach for analyzing the fundamental physics of ionic motion
in a turbulent magnetic field treats the total magnetic field $\vec B$ 
as a spatially fluctuating
component $\delta{\vec B}$ superimposed onto a static
background component ${\vec  B_0}$, where $\delta \vec B$ is
generated by summing over a large number
of  randomly polarized  transverse waves with wavelengths
$\lambda_n = 2\pi/k_n$, logarithmically spaced
between $\lambda_{min}$ and $\lambda_{max}$
(e.g., Giancoli \& Jokipii 1994; Casse et al. 2006; O'Sullivan et al. 2009;
Fatuzzo et al. 2010).
Adopting a static turbulent field removes the necessity of specifying 
a dispersion relation between the wavevectors $k_n$ and their
corresponding angular frequencies $\omega_n$.  This approach therefore
allows one to consider highly non-linear turbulence ($\delta B >> B_0$), or even remove 
the background component altogether. Of course, turbulent magnetic fields in cosmic 
environments are not static. Nevertheless, a static formalism in spatial 
diffusion calculations of relativistic particles is justified for environments 
in which the Alfv\'en speed is much smaller than the speed of light.
 
This paper focuses on the energy diffusion of cosmic rays propagating
through a turbulent magnetic environment, which then requires 
the use of a time-dependent formalism in order to self-consistently include the
fluctuating electric fields that must also be present. 
Toward that end,  we assume that the GC environment
is well represented by a nonviscous, 
perfectly conducting fluid threaded by a uniform static field $\vec B_0 = B_0 \hat z$, 
and use linear MHD theory to guide us. 
In general, three types of MHD waves exist in the linear regime---Alfv\'en, fast and slow.  
Following the formalism of O'Sullivan et al. (2009), we consider
here only Alfv\'en waves, so that the turbulent magnetic field 
is defined by the sum of $N$ randomly directed waves
\be
 \delta \vec B = \sum_{n=1}^N \vec A_n \,  e^{i( \vec k_n \cdot \vec r-\omega_n t+\beta_n)}\,,
\ee
where the direction of each propagation vector $\vec k_n$ is set through a random choice of polar angles
$\theta_n$ and $\phi_n$, and the phase of each term is set through a random choice
of $\beta_n$.

Alfv\'en waves don't compress the fluid through which they propagate, 
and are therefore characterized by
a fluid velocity $\vec v$ that satisfies the condition $\vec k \cdot \vec v = 0$.
This condition in turn implies that $\vec v \cdot \hat B_0 = 0$.
We can therefore write the fluid velocity associated with the $n$-th term
in Equation (1) as
\be
\delta \vec v_n = \pm A_n \, {v_A\over B_0} \, {\hat z \times \vec k_n \over | \hat z \times \vec k_n |}
 \,  e^{i( \vec k_n \cdot \vec r-\omega_n t+\beta_n)}\,,
 \ee
where the sign is chosen randomly for each term in the sum.  The dispersion
relation for Alfve\'nic waves is given by the expression  
\be
\omega_n = v_A k_n |\cos\theta_n| \,,
\ee
where $v_A$ is the Alfv\'en speed, and
$\theta_n$ is the angle between ${\vec k_n}$ and ${\vec B_0}$

The corresponding magnetic field for each wave then follows from
the linear form of Amp\`ere's Law, as given by 
\be
\vec A_n = \mp A_n \, {\vec k_n \cdot \hat z \over | \vec k_n \cdot \hat z |} \, {\hat z \times \vec k_n \over | \hat z \times \vec k_n |}
 \,.  
 \ee
This formalism is identical to that of O'Sullivan et al. (2009),
with the exception that we allow for a random choice of sign
in each term.  We find, however, that including a random sign has no effect on
the statistical output measures of our numerical simulations.   

The total number of terms in the sum is given by
$N = N_k \log_{10}[k_{max}/k_{min}]$, and the values of
$k_n$ are evenly spaced on a logarithmic scale between $k_{min} = 2\pi / \lambda_{max}$ and
$k_{max} = 2\pi/\lambda_{min}$.  
The desired spectrum of the turbulent magnetic field is set
through the appropriate choice of $\Gamma$ in the scaling
\begin{equation}
A_n^2 = A_1 ^2\left[{k_n \over k_1} \right]^{-\Gamma}
{\Delta k_n\over \Delta k_1} 
= A_1^2\left[{k_n \over k_1} \right]^{-\Gamma+1}
\end{equation}
(e.g., $\Gamma = 3/2$ for Kraichnan and 5/3 for Kolmogorov turbulence).
The value of $A_1$ is set by a parameter $\xi$ that specifies the averaged energy 
density of the turbulent field via the definition 
\begin{eqnarray}
\langle \delta B^2 \rangle = \langle \sum_n \sum_{n'}  \delta \vec B_n 
\cdot \delta \vec B_{n'}^* \rangle \nonumber \\ 
 = A_1^2 \sum_n \left[{k_n \over k_1}\right]^{-\Gamma+1}\,
 = \xi\, B_0^2\;,
\end{eqnarray}
where the $n \ne n'$ terms average to zero.   
We note that for our adopted scheme, the value of $\Delta k_n / k_n$
is the same for all values of $n$.
We further note that $\xi = 2$ corresponds to the real part of the turbulent field
having the same energy density as a uniform field $B_0$.

Naively extending the results from linear MHD theory 
to our formalism, one would obtain the total electric field $\delta {\vec E}$ 
associated with the turbulent 
magnetic field defined by Eq. (1)  by summing over the terms
\be
 \delta{\vec E_n}=  -\delta \vec v_n \times B_0\,.
\ee
Although $ \delta{\vec E} \cdot  {\vec B_0}$ = 0, 
the second order term $\delta {\vec E}\cdot \delta {\vec B} \ne 0$.   The presence of
an electric field component parallel to the magnetic field in 
this second order term can significantly increase the acceleration efficiency artificially, 
especially if the formalism is extended to the nonlinear regime 
($\delta B \sim$$B_0$).   However, the interstellar medium is highly conductive, and as such, 
any electric field component parallel to the magnetic field should be quickly
quenched.  
To circumvent this problem, we
adopt the formalism of O'Sullivan et al. (2009) and 
first obtain the total fluid velocity $\delta {\vec v}$ via the summation
\be
\delta {\vec v} = \sum_{n=1}^N  \delta \vec v_n\,.
\ee
We then use the MHD condition to set the total electric field:
\be
\delta {\vec E} = - {\delta {\vec v}\over c} \times\vec B\;,
\ee
where $\vec B = B_0\hat z + \delta\vec B$.

\section{Numerical Analysis}

The equations that govern the motion of a relativistic charged proton
with Lorentz factor $\gamma$
through the turbulent medium are
\be
{d\over  dt}  (\gamma m_p \vec v)=  e \left[\delta {\vec E} +
{\vec v \times \vec B\over c}\right]\;,
\ee
and
\be
{d {\vec r}\over dt} = \vec v\;.
\ee
Although these equations are deterministic, the chaotic nature of motion through turbulent fields 
necessitates a statistical analysis.  

We define a single experiment as a numerical investigation
of particle dynamics through a given environment
and a given particle injection energy.
The environment is specified by the underlying field strength $B_0$
and Alfv\'en speed  $v_A$ (see \S2), and the turbulent
fields are specified by the parameters $\Gamma$, $\lambda_{max}$, 
$\lambda_{min}$, $N_k$ and $\xi$.  For each experiment, we
numerically integrate the
equations of motion for
$N_p = 1000$ protons randomly injected  from the origin 
with the same initial energy, as specified by the Lorentz factor
$\gamma_0$.
The equations of motion 
are integrated for a time $\Delta t = 100 \lambda_{max}/c$, with
each particle sampling its own unique magnetic field structure
(i.e., the values of $\beta_n$, $\theta_n$, 
$\phi_n$ and the choice of a $\pm$ are 
chosen randomly for each particle). 

As is well known, the diffusion coefficients provide a useful output measure
for the characterization of the diffusion process since their values are constant
once the particles are in the diffusion regime.   We therefore adopt
$D_\gamma \equiv \langle \Delta \gamma^2 \rangle / (2 \Delta t)$,
$D_\perp \equiv \langle \Delta x^2 \rangle / (2 \Delta t)$,
and $D_{||} \equiv \langle \Delta z^2 \rangle / (2 \Delta t)$ as the output
measures of our experiments (recall that the underlying field $\vec B_0$ is in the
$\hat z$ direction).  The diffusion constants can then be used to obtain the 
``acceleration time" $\tau_{acc} = \gamma_0^2 / D_\gamma$, which characterizes
how long it would take low energy particles to diffuse to $\gamma_0 m_p c^2$ energies
(and hence, attain 
a high energy tail $\ga \gamma_0 m_p c^2$), and the escape time
$\tau_{esc} = R^2/ D_{||}$, which characterizes how long it would take those
particles to diffuse a distance $R$ along the 
preferential $\hat z$
direction.   

As elaborated upon in \S 3 below, a given experiment is defined by the
parameters $B_0$, $v_A$, $\Gamma$, $\lambda_{max}$, $\xi$, and $\gamma_0$, 
and the output measures are the diffusion coefficients $D_\gamma$,
$D_\perp$ and $D_{||}$.  These values are summarized in Table 1 for
all experiments performed in this work.

\subsection{Baseline experiment}
Our first goal is to find a numerical scheme
that minimizes computer time 
without sacrificing accuracy.  
Toward that end, we perform our first numerical
experiment for $\gamma_0 = 10^6$ particles 
injected into the intercloud
environment ($B_0 = 50 \,\mu$G, $v_A = 35$ km/s).  We adopt a 
baseline set of turbulent field parameters $\Gamma = 5/3$,
$\lambda_{max} = 1$ pc, $\lambda_{min} = 0.002$ pc, $\xi = 2$
and $N_k = 25$.   We note that the particle radius of gyration,
\be
R_{g0} = 0.02\, {\rm pc} \left({\gamma_0\over 10^6} \right)
\left({B_0\over 50 \,\mu{\rm G}}\right)^{-1}\;,
\ee
falls comfortably within the values of $\lambda_{min}$ and $\lambda_{max}$.

As can be seen from Figure 1, a single particle's energy,
as characterized by $\Delta\gamma / \gamma_0 = \gamma / \gamma_0 - 1$,  changes in a
random-like fashion.  As such, the energy distribution 
for an ensemble of particles injected with the same energy
becomes normal once the particles have fully sampled the
turbulent nature of the accelerating electric fields.  This
point is clearly illustrated by Figure 2, which shows the 
distribution of $\Delta\gamma$ values at time $t = 100 \lambda_{max}/c$
for the $1000$ particles tracked in our baseline experiment (Experiment 1).  One can  
therefore quantify the stochastic acceleration 
of particles in turbulent fields through the variance of the resulting
distributions of initially mono-energetic particles.  To illustrate this point, we
plot in Figure 3 the variance $\sigma_\gamma\equiv\sqrt{\langle\Delta\gamma^2\rangle}$ of the
$\Delta\gamma$ distribution as a 
function of time for the particles tracked
in our baseline experiment.
As expected from the random nature of stochastic
acceleration, $\sigma_\gamma \propto \sqrt{t}$ once particles have had
a change to sample the turbulent nature of the underlying fields, i.e., for
$t \ga \lambda_{max}/c$.

\begin{figure}
\figurenum{1}
\begin{center}
{\centerline{\epsscale{0.90} \plotone{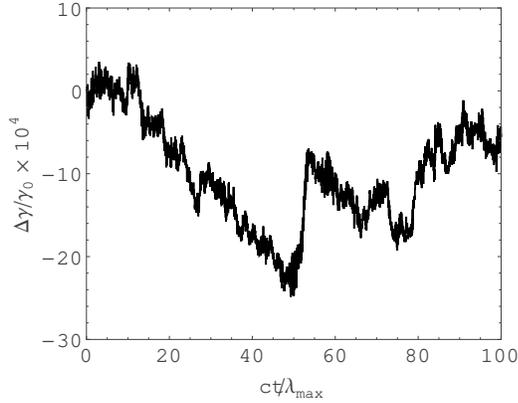} }}
\end{center}
\figcaption{The fractional change in particle energy
$\Delta\gamma / \gamma_0$
as a function of time for a $\gamma_0 = 10^6$ particle
injected into an intercloud-like environment  
 ($B_0 = 50\,\mu$G, $v_A = 35$ km/s)
 with an Alfv\'enic turbulent field defined by the parameters
$\lambda_{max} = 1$ pc, $\lambda_{min} = 0.002$ pc, 
$\Gamma = 3/2$, $\xi = 2.0$, and $N_k = 25$.  
}
\end{figure}

\begin{figure}
\figurenum{2}
\begin{center}
{\centerline{\epsscale{0.90} \plotone{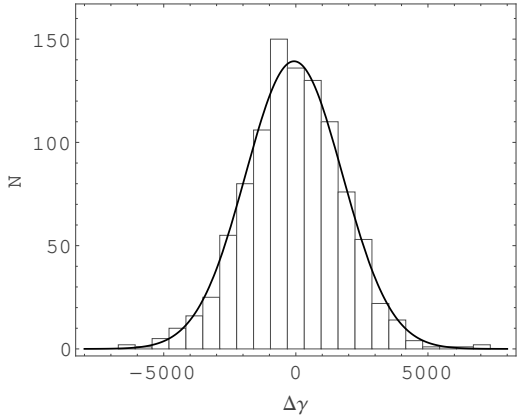} }}
\end{center}
\figcaption{The particle energy distribution for the ensemble of
$N_p = 1,000$ particles at time $t = 100 \lambda_{max}/c$ in our baseline numerical experiment (Experiment 1).
The solid line shows a Gaussian fit to the data.
}
\end{figure}

\begin{figure}
\figurenum{3}
\begin{center}
{\centerline{\epsscale{0.90} \plotone{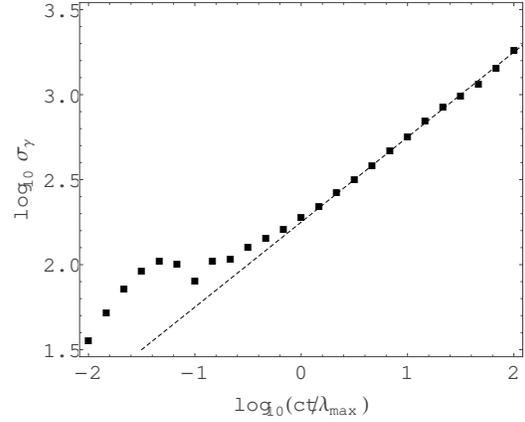} }}
\end{center}
\figcaption{The variance of the particle energy distribution as a function 
of time for the $N_p = 1,000$ particles in our baseline numerical experiment
(Experiment 1).
The dotted line serves as a reference and has a slope of $1/2$, clearly
indicating that $\sigma_\gamma\propto\sqrt{t}$ for time $t \ga \lambda_{max}/c$.
}
\end{figure}

The spatial diffusion of particles will be different in the
parallel and perpendicular  directions to the underlying
magnetic field $B_0 \hat z$, resulting in different
variances in the distributions of particle displacement along
and across the underlying field.   To illustrate this point, we
plot $\sigma_x \equiv \sqrt{\langle \Delta x^2\rangle}$ and
$\sigma_z \equiv\sqrt{ \langle \Delta z^2\rangle}$ as a function of
time in Figure 4 for the particles in our baseline experiment.  As expected,
particles diffuse farther in the direction parallel to the magnetic field, 
and  both output measures become $\propto
\sqrt{t}$ at time $t \ga \lambda_{max}/c$.

\begin{figure}
\figurenum{4}
\begin{center}
{\centerline{\epsscale{0.90} \plotone{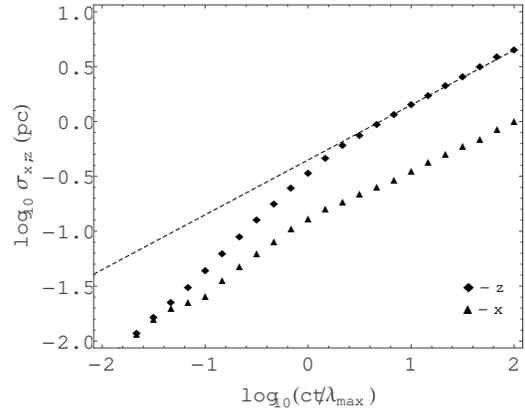} }}
\end{center}
\figcaption{The variance of the particle spatial distributions
perpendicular ($x$) and parallel ($z$) 
to the underlying field $\vec B_0$ as a function 
of time for the $N_p = 1,000$ particles in our baseline numerical experiment
(Experiment 1).
The dotted line serves as a reference and has a slope of $1/2$, clearly
indicating that $\sigma_x$ and $\sigma_z$ are proportional to $\sqrt{t}$ for time $t \ga \lambda_{max}/c$.
}
\end{figure}

A fundamental issue in this analysis is what value of $N_k$
will allow our discrete treatment of the turbulent field to
adequately represent the continuous fields found in nature.  Toward that end,
we repeated our baseline experiment  with $N_k$ = 250.  
The difference in output measures were smaller 
than 10\%, indicating that setting $N_k = 25$ provides good accuracy in our
results.
We also repeated our baseline experiment with $\lambda_{min} = 0.0002$ pc.  
Again, the difference in output measures were smaller 
than 10\%, indicating that
our output measures are not sensitive to the value of $\lambda_{min}$ 
so long as the particle radius of gyration $R_g>\lambda_{min}$ (see also Fatuzzo et al. 2010).

Guided by the results of our baseline analysis, we will perform the remainder of our
experiments using $N_k = 25$ and $\lambda_{min} =
0.1 R_{g0}$, where
\be
R_{g0} = {\gamma_0 m_p c^2\over e B_0}\;,
\ee
is the radius of gyration for the injected particles in the absence
of turbulence.  The governing equations for each particle will be 
integrated out to a time sufficiently long to ensure that $\sigma_\gamma$,
$\sigma_x$ and $\sigma_z$ are proportional to $\sqrt{t}$
by the end of the
integration (usually $\Delta t = 100 \lambda_{max}/c$), 
and the particles have therefore fully sampled the turbulent
nature of the magnetic and electric fields.  This scheme is expected to
provide accurate results with minimal computing resources. 

As a consistency check, we perform a set of experiments (2 -- 7) using 
the same physical conditions corresponding to models W3 
and X3 presented in O'Sullivan et al. (2009;  see Table 1 and
Figures 3 and 4).  Specifically, we consider
a physical environment defined by the parameters $B_0 = 3 \,\mu$G
and $v_A = 0.002$ c, and a turbulent field defined by $\Gamma = 5/3$
(Kolmogorov) and $\lambda_{max} = 1$ kpc for $\xi = 0.2$ (corresponding to
model W3 for which $\langle\delta B/B_0\rangle^2 = 0.1$) and 
$\xi = 2$ (corresponding to
model X3 for which $\langle\delta B/B_0\rangle^2 = 1$).
We compare our results for the
acceleration time $\tau_{acc}$ to those obtained
by O'Sullivan et al. (2009) in Figure 5.  
While there is general agreement between these results, our values
of the 
acceleration times are consistently about a factor of two greater than those
to which we are comparing.  

As a final consistency check, we compare in Figure 6 the spatial diffusion coefficients
$D_\perp$ and $D_{||}$ obtained for Exp. 5 -- 7 to those obtained 
by Fatuzzo et al. (2010) under the same physical conditions.  We note that the turbulent magnetic
field used in our earlier work is of the form given by Equation 1, but
with 
\be
\vec A_n = A_n \left(\cos\alpha_n \; \hat y' \pm i \sin\alpha_n \;\hat z'\right)\;,
\ee
where the $y' - z'$ plane is normal to the $\vec k$ direction and
$\alpha_n$ is picked at random for each $n$ term. 
As can be seen from Figure 6, there is good agreement between results,
indicating that the turbulent magnetic field adopted in this
work is equivalent to that used in our earlier work.  We note also that
the presence of a weak turbulent electric field (i.e., $\delta E << \delta B$), which 
was not included in the calculations of Fatuzzo et al. (2010),
has a negligible effect on the spatial diffusion of particles.

\begin{figure}
\figurenum{5}
\begin{center}
{\centerline{\epsscale{0.90} \plotone{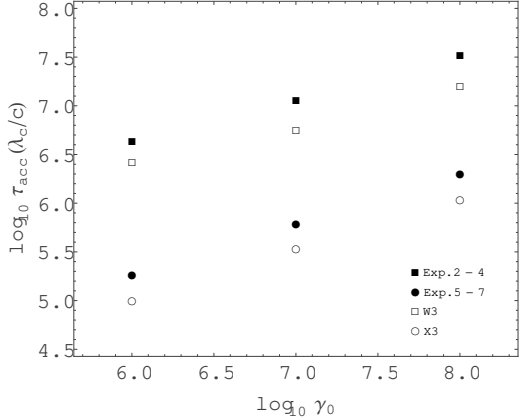} }}
\end{center}
\figcaption{Acceleration time $\tau_{acc}$ as a function of the injected
particle Lorentz factor, for both our work and the work of O'Sullivan et al. (2009).
Open squares (W3) and circles (X3) denote the results of O'Sullivan et al. (2009).  Solid squares
and circles denote our results. We note that $\lambda_c = 0.77 \lambda_{max} / (2\pi)$.}
\end{figure}

\begin{figure}
\figurenum{6}
\begin{center}
{\centerline{\epsscale{0.90} \plotone{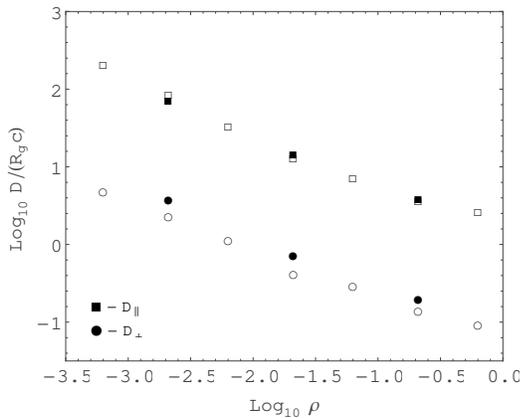} }}
\end{center}
\figcaption{Diffusion coefficients $D_\perp$ and $D_{||}$ as a function of rigidity
$\rho = 2\pi R_{g0}/\lambda_{max}$ for experiments 5 -- 7 [solid squares and circles]
compared with results obtained by Fatuzzo et al. (2010) for which $\xi = 2$ (open
squares and circles).}
\end{figure}

\subsection{Survey of Parameter Space}
As noted above,  the spatial and energy
diffusion of particles through turbulent fields is not
sensitive to the minimum wavelength so 
long as the particle's radius of gyration exceeds $\lambda_{min}$.
For a given environment, as defined by $B_0$ and $v_A$, 
the diffusion process is thus dependent upon 
the maximum turbulence wavelength $\lambda_{max}$,
the turbulent field strength, as characterized by $\xi$,
and the turbulence spectrum, as characterized by the spectral index $\Gamma$.

Quasi-linear theory predicts that the energy diffusion
coefficient for relativistic particles scales as
\be
D_\gamma \approx {v_A^2\over c^2} \,{\delta B^2\over B_0^2}
\left({R_g\over \lambda_{max}}\right)^{\Gamma-1} 
{p^2c^2\over R_g c} \,,
\ee
where $p$ is the particle momentum (Schlickeiser 1989).  
For relativistic particles, the energy diffusion coefficient 
should therefore scale to our model parameters as
\be
D_\gamma \propto \xi\, \lambda_{max}^{1-\Gamma} \, \gamma_0^\Gamma\,.
\ee
Equation (15) however does not appear to be valid
in the strong turbulence limit (O'Sullivan et al. 2009).  We therefore investigate
how the energy diffusion coefficient depends upon $\lambda_{max}$
and $\xi$ for Kolmogorov ($\Gamma = 5/3$) turbulence.  
Specifically, we perform two sets of experiments designed to
explore parameter space around our baseline values. 
For the first set (Experiments 8 -- 10), we vary 
$\lambda_{max}$ from 0.32 -- 10 pc, while in   
the second set (Experiments 11 -- 13) we vary $\xi$ between 0.2 -- 6.4.

The energy diffusion coefficients for Experiments 1 and 8 -- 10
are presented in Figure 7, with the best line fit to the data 
indicating that $D_\gamma \propto
\lambda_{max}^{-0.47}$.  This results confirms that
quasi-linear theory, which predicts that
$D_\gamma \propto \lambda_{max}^{-0.67}$ for
Kolmogorov turbulence, is not applicable in the strong turbulence ($\xi \ga 1$)
limit. 
The energy diffusion coefficients for Experiments 1 and 8 -- 10
are presented in Figure 8, with the best line fit to the data 
indicating that $D_\gamma \propto
\xi^{1.2}$ in both the weak and strong turbulence limit.

\begin{figure}
\figurenum{7}
\begin{center}
{\centerline{\epsscale{0.90} \plotone{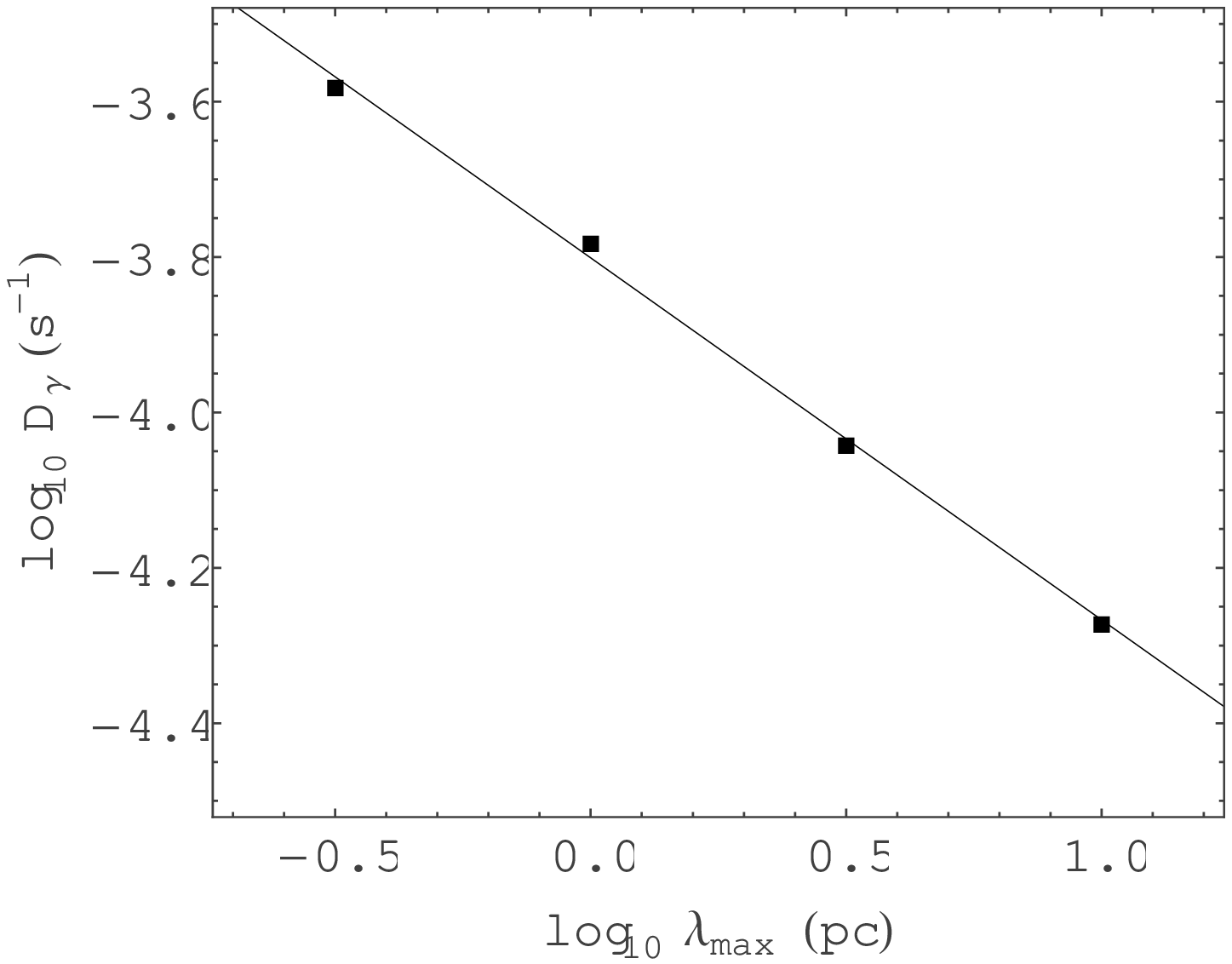} }}
\end{center}
\figcaption{The energy diffusion coefficients $D_\gamma$ as a function of maximum
turbulence wavelength 
$\lambda_{max}$  for experiments 1 and 8 -- 10.  The best line fit to the data has a slope
of $-0.47$. }
\end{figure}

\begin{figure}
\figurenum{8}
\begin{center}
{\centerline{\epsscale{0.90} \plotone{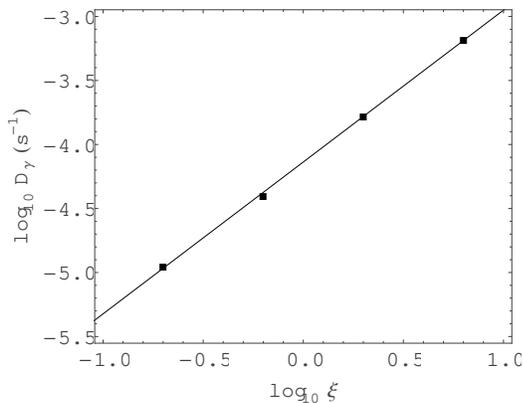} }}
\end{center}
\figcaption{The energy diffusion coefficients $D_\gamma$ as a function of $\xi$
 for experiments 1 and 11 -- 14. The best line fit to the data has a slope
of $1.2$. }
\end{figure}

\section{Application to the GC Environment}
The stochastic acceleration of protons in a magnetically turbulent environment
is essentially a 1-D random walk process that results in a 
particle energy distribution that broadens in time.  
Our goal in this section is to calculate the characteristic particle energy
for both the inter-cloud medium and the molecular cloud
region observed at the GC.
As shown in \S 4, a numerical treatment of particle diffusion
requires an integration time-step $\delta t \sim$$0.1 \lambda_{min}
\sim$$0.01 R_g$, but a total integration time that is $\Delta t \sim$$100 \lambda_{max}$.
Numerically integrating the equations of motion
for protons that are energized from a thermal state to TeV 
energies is therefore not computationally feasible given the small
radius of gyration of thermal particles. 

As such, we determine the characteristic particle energy in each region by comparing the 
``acceleration time" $\tau_{acc} = \gamma_0^2 / D_\gamma$, which characterizes
how long it would take a distribution of low-energy particles to attain 
energies $\sim$$\gamma_0 m_p c^2$, with the escape time
$\tau_{esc} = R^2/ D_{||}$, which characterizes how long it would take those
particles to diffuse a distance $R$ along the 
preferential $\hat z$
direction, and the proton cooling time $\tau_{pp}$, which characterizes
how long a proton can move through a region before losing a significant
fraction of its energy to $pp$ scattering with the ambient medium.

To determine how the acceleration and escape times depend on particle energy
for the
inter-cloud medium ($B_0 = 50\,\mu$G, $v_A = 35$ km/s, $R = 200$ pc), 
we perform a set of Experiments (14 -- 18) using our base-line parameters
$\Gamma = 5/3$, $\lambda_{max} = 1$ pc, and $\xi = 2$. 
We perform two additional experiments (19 \& 20) at $\gamma_0 = 10^5$
and $\gamma_0 = 10^6$ for the same environment, but with $\Gamma = 3/2$.
We perform a final set of Experiments (21 -- 26) to
determine how the acceleration and escape times depend on particle energy
for the molecular cloud medium ($B_0 = 500\,\mu$G, $v_A = 11$ km/s, $R = 30$ pc).

The energy loss rate of protons with energies needed to produce 
$\pi^0$-decay $\gamma$-rays is dominated by nuclear energy losses due to
$pp$ scattering with the ambient medium (Aharonian \& Atoyan 1996). As such, the 
cooling time, $\tau_{pp}$, of the protons depends on the pp-scattering cross-section, $\sigma_{pp}$, and 
the inelasticity parameter, $\kappa$. Over a broad range of proton energies, neither of these quantities 
significantly varies so the usual method is to adopt the constant average values $\sigma_{pp} \approx$ 
40 mb and $\kappa \approx$ 0.45 (see, e.g., Markoff et al. 1997). That being the case, the proton 
cooling time becomes independent of proton energy:
\begin{equation}
\tau_{pp} = (n_H c \kappa \sigma_{pp})^{-1} \;,
\end{equation}
where $n_H$ is the number density of ambient protons.
The value of $\tau_{pp}$ is therefore $\approx 2 \times 10^{14}$ s 
for the inter-cloud region and $\approx 2 \times 10^{11}$ within the molecular cloud region.

Our principal results are presented in Figures 9 and 10. A characteristic 
particle energy for each region can be estimated by the crossing point 
where $\tau_{acc}$ becomes greater than either  $\tau_{esc}$ or $\tau_{pp}$.
Interestingly, this value is $\sim$$3$ Tev for the intercloud region, indicating
that the TeV cosmic-rays associated with the HESS observations can thus
be produced via stochastic acceleration within the turbulent intercloud environment. 
This result appears to hold true for both Kolmogorov ($\Gamma = 5/3$) and
Kraichnan ($\Gamma = 3/2$) turbulence, and suggests that energy diffusion
in the strong turbulence limit is not sensitive to the turbulence power-spectrum.
In contrast, while the acceleration time associated with the molecular clouds
are comparable to those in the intercloud medium, the escape time and the 
proton cooling time are significantly shorter.  As such, it is clear from Figure 10
that TeV protons cannot
be energized within the molecular cloud environment.

\begin{figure}
\figurenum{9}
\begin{center}
{\centerline{\epsscale{0.90} \plotone{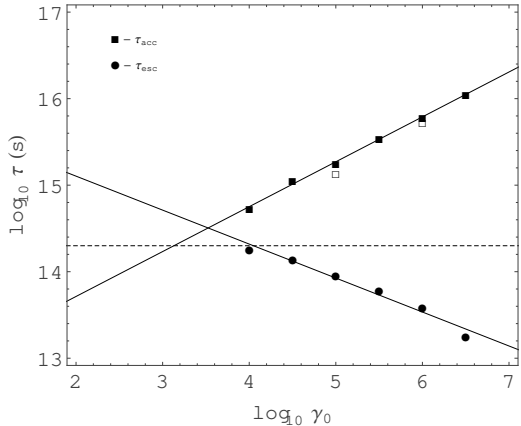} }}
\end{center}
\figcaption{The acceleration time $\tau_{acc}$ and escape time $\tau_{esc}$
for the intercloud environment as a function of injection energy 
(as characterized by the Lorentz factor $\gamma_0$) for experiments 
14 -- 20 (solid squares and circles) and experiments 19 -- 20
(open squares).  The solid lines represent the best-line fits to the data from 
experiments 14 -- 20,
and the dashed line represents the value of $\tau_{pp}$ in the intercloud region
.}
\end{figure}

\begin{figure}
\figurenum{10}
\begin{center}
{\centerline{\epsscale{0.90} \plotone{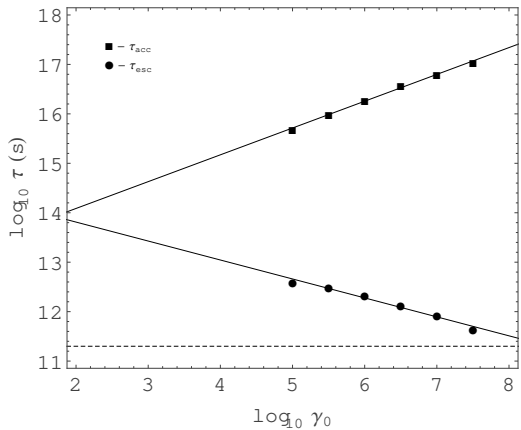} }}
\end{center}
\figcaption{The acceleration time $\tau_{acc}$ and escape time $\tau_{esc}$
for the molecular cloud environment as a function of injection energy 
(as characterized by the Lorentz factor $\gamma_0$)  
for experiments 21 --  26.  The solid lines represent the best-line fits to the data,
and the dashed line represents the value of $\tau_{pp}$ in the GC molecular clouds.}
\end{figure}

\section{Conclusion}
The purpose of this paper has been to assess the 
feasibility of stochastic acceleration
within the GC region in producing the TeV cosmic-rays 
revealed by the diffuse HESS emission correlated with the 
molecular gas distributed along the GC ridge.
As shown in previous work, this emission can be produced by
cosmic rays with an energy distribution that has a high-energy tail
extending out beyond a few TeV.  Such a distribution is
naturally produced by the stochastic acceleration of particles 
in a magnetically turbulent environment (which is effectively a 1-D random
walk in energy). Thus, the remaining question is whether or not this mechanism is
efficient enough to energize protons to the required TeV energies in 
the highly conductive interstellar medium permeating the galactic center.

To resolve this issue, we performed a series of numerical experiments 
in order to calculate the spatial and energy diffusion coefficients for
protons in both the molecular cloud and the intercloud medium located within 
the inner several hundred parsecs of the galaxy.  
We then estimated the characteristic energy of the resulting particle distribution by comparing the 
acceleration time required for particles to reach a certain energy, the escape time
which characterizes how long it takes protons to diffuse out of each region, and the cooling time which characterizes
how long a proton can move through a region before losing a significant
fraction of its energy to $pp$ scattering with the ambient medium.

Our results indicate that for the physical conditions observed in the intercloud medium,
together with reasonable estimates of the turbulent field strength and maximum turbulence
wavelength, protons can be accelerated to characteristic energies $\sim$$3$ TeV, 
indicating that the TeV cosmic-rays observed by HESS can thus
be produced via stochastic acceleration within the turbulent intercloud environment. 
In contrast, while the acceleration time associated with the molecular clouds
are comparable to those in the intercloud medium, the escape time and the 
proton cooling time are significantly shorter.  As such, protons cannot be energized
to TeV energies within the molecular cloud environment.  

These results are very encouraging, not only because the idea of
stochastic acceleration in a turbulent magnetic field at the GC
appears to be a viable mechanism for producing the cosmic rays
observed in that region, but especially because the characteristic
energy attained by the relativistic protons matches the observations
very well. We note, in this regard, that the physical conditions used in
our simulations are unique to the GC. As such, cosmic rays like those 
observed by HESS are not easily produced by this mechanism anywhere else
in the Galaxy.

So the investigation now turns to the very important subsequent question,
which we already alluded to in the introduction, viz. can we understand
the origin of cosmic rays observed at Earth, at least up to energies
$\sim$3--5 TeV, as the result of stochastic acceleration at the GC
followed by energy-dependent diffusion and escape across the Milky Way? 
Simulations designed to address this issue are currently underway, and the 
results will be reported elsewhere.

\acknowledgments

This work was supported by Xavier University through the Hauck Foundation.
At the University of Arizona, partial support was provided by NASA grant
NNX08AX34G and ONR grant N00014-09-C-0032. In addition, FM is grateful to 
Amherst College for its support through a John Woodruff Simpson Lectureship. 
Finally, we are happy to acknowledge Stephen O'Sullivan for helpful discussions.






\begin{deluxetable}{cccccccccc}
\tablecolumns{10}
\tablewidth{0pc}
\tablecaption{Summary of Experiments}
\tablehead{
\colhead{Exp} & \colhead{$B_0$ ($\mu$G)}   & \colhead{$v_A$ (km/s)}    
& \colhead{$\Gamma$} 
& \colhead{$\lambda_{\rm max}$}  (pc) & \colhead{$\xi$} &  \colhead{$\gamma_0$}  
 &\colhead{$D_\gamma$ (s$^{-1}$)} &\colhead{$D_\perp$ (cm$^2$ s$^{-1}$)}
&\colhead{$D_{||}$ (cm$^2$ s$^{-1}$)} 
 }
\startdata
1 & 50  & $35$ &$5/3$& 1 & 2& $10^6$ & $1.6\times 10^{-4}$ & $4.9\times 10^{26}$ &$9.8\times 10^{27}$ \\
2 & 3  & $600$ &$5/3$& $10^3$ & 0.2& $10^8$ & $2.4\times 10^{-2}$ & $1.4\times 10^{29}$ &$2.6\times 10^{32}$ \\
3 & 3  & $600$ &$5/3$& $10^3$ & 0.2& $10^7$ & $6.8\times 10^{-4}$ & $9.4\times 10^{28}$ &$1.2\times 10^{32}$ \\
4 & 3  & $600$ &$5/3$& $10^3$ & 0.2& $10^6$ & $1.8\times 10^{-5}$ & $7.0\times 10^{28}$ &$6.6\times 10^{31}$ \\
5 & 3  & $600$ &$5/3$& $10^3$ & 2& $10^8$ & $3.9\times 10^{-1}$ & $6.2\times 10^{29}$ &$1.2\times 10^{31}$ \\
6 & 3  & $600$ &$5/3$& $10^3$ & 2& $10^7$ & $1.3\times 10^{-2}$ & $2.3\times 10^{29}$ &$4.6\times 10^{30}$ \\
7 & 3  & $600$ &$5/3$& $10^3$ & 2& $10^6$ & $4.2\times 10^{-4}$ & $1.2\times 10^{29}$ &$2.3\times 10^{30}$ \\
8 & 50  & $35$ &$5/3$& 0.32 & 2& $10^6$ & $2.6\times 10^{-4}$ & $2.6\times 10^{26}$ &$6.7\times 10^{27}$ \\
9 & 50  & $35$ &$5/3$& 3.2 & 2& $10^6$ & $9.2\times 10^{-5}$ & $9.3\times 10^{26}$ &$1.8\times 10^{28}$ \\
10 & 50  & $35$ &$5/3$& 10 & 2& $10^6$ & $5.4\times 10^{-5}$ & $1.7\times 10^{27}$ &$4.2\times 10^{28}$ \\
11 & 50  & $35$ &$5/3$& 1 & 0.2& $10^6$ & $1.1\times 10^{-5}$ & $1.4\times 10^{26}$ &$2.1\times 10^{29}$ \\
12 & 50  & $35$ &$5/3$& 1 & 0.64& $10^6$ & $4.0\times 10^{-5}$ & $2.9\times 10^{26}$ &$4.6\times 10^{28}$ \\
13 & 50  & $35$ &$5/3$& 1 & 6.4& $10^6$ & $6.6\times 10^{-4}$ & $6.9\times 10^{26}$ &$2.6\times 10^{27}$ \\
14 & 50  & $35$ &$5/3$& 1 & 2& $3.2 \times 10^6$ & $8.7\times 10^{-4}$ & $8.4\times 10^{26}$ &$2.1\times 10^{28}$ \\
15 & 50  & $35$ &$5/3$& 1 & 2& $3.2\times 10^5$ & $2.9\times 10^{-5}$ & $2.9\times 10^{26}$ &$6.3\times 10^{27}$ \\
16 & 50  & $35$ &$5/3$& 1 & 2& $10^5$ & $5.6\times 10^{-6}$ & $2.0\times 10^{26}$ &$4.3\times 10^{27}$ \\
17 & 50  & $35$ &$5/3$& 1 & 2& $3.2\times 10^4$ & $1.0\times 10^{-6}$ & $1.3\times 10^{26}$ &$2.8\times 10^{27}$ \\
18 & 50  & $35$ &$5/3$& 1 & 2& $10^4$ & $1.9\times 10^{-7}$ & $9.6\times 10^{25}$ &$2.1\times 10^{27}$ \\
19 & 50  & $35$ &$3/2$& 1 & 2& $10^6$ & $1.9\times 10^{-4}$ & $3.9\times 10^{26}$ &$9.6\times 10^{27}$ \\
20 & 50  & $35$ &$3/2$& 1 & 2& $10^5$ & $7.5\times 10^{-6}$ & $1.3\times 10^{26}$ &$2.3\times 10^{27}$ \\
21 & 500  & $11$ &$5/3$& 1 & 2& $3.2\times 10^7$ & $9.1\times 10^{-3}$ & $8.6\times 10^{26}$ &$2.0\times 10^{28}$ \\
22 & 500  & $11$ &$5/3$& 1 & 2& $10^7$ & $1.6\times 10^{-3}$ & $4.5\times 10^{26}$ &$1.0\times 10^{28}$ \\
23 & 500  & $11$ &$5/3$& 1 & 2& $3.2\times 10^6$ & $2.7\times 10^{-4}$ & $2.8\times 10^{26}$ &$6.3\times 10^{27}$ \\
24 & 500  & $11$ &$5/3$& 1 & 2& $10^6$ & $5.5\times 10^{-5}$ & $2.0\times 10^{26}$ &$4.0\times 10^{27}$ \\
25 & 500  & $11$ &$5/3$& 1 & 2& $3.2\times 10^5$ & $1.0\times 10^{-5}$ & $1.4\times 10^{26}$ &$2.8\times 10^{27}$ \\
26 & 500  & $11$ &$5/3$& 1 & 2& $10^5$ & $2.1\times 10^{-6}$ & $1.1\times 10^{26}$ &$2.2\times 10^{27}$ \\

\enddata
\end{deluxetable}

\end{document}